# Electron beam characterization with beam loss monitors

L. Verra[,1,2,3,*] M. Turner,[1] S. Gessner,[1] E. Gschwendtner,[1] F. M. Velotti,[1] and P. Muggli[2]

[1]*CERN, Geneva 1211, Switzerland*
[2]*Max-Planck Institute for Physics, Munich 80805, Germany*
[3]*Technical University Munich, Munich 85748, Germany*

We present a method to measure the transverse size and position of an electron or proton beam, close to the injection point in plasma wakefields, where other diagnostics are not available. We show that transverse size measurements are in agreement with values expected from the beam optics with a < 10% uncertainty. We confirm the deflection of the low-energy (∼18 MeV) electron beam trajectory by Earth's magnetic field. This measurement can be used to correct for this effect and set proper electron bunch injection parameters. The advanced wakefield experiment at CERN (AWAKE) relies on these measurements for optimizing electron injection.

## I. INTRODUCTION

### A. The AWAKE experiment

AWAKE [1], the advanced wakefield experiment at CERN, recently demonstrated acceleration of externally injected electrons in plasma wakefields resonantly excited by a self-modulated [2,3] relativistic proton bunch [4].

The core of the experiment is a 10-m-long rubidium vapor source [5]: a long, fluid-heated heat exchanger evaporates rubidium at 180 °C–230 °C to reach the required vapor density of $0.5$–$10 \times 10^{14}$ atoms/cm$^3$. A 120 fs, <450 mJ laser pulse ($\lambda = 780$ nm) ionizes the rubidium vapor, creating a plasma cylinder with a radius of approximately 1 mm [6]. The vapor source is connected to the beam line at each end by a 10-mm-diameter aperture. The 400 GeV/c proton bunch provided by the CERN Super Proton Synchrotron (SPS) and delivered by a 750-m-long transfer line [7] with $3 \times 10^{11}$ particles drives the plasma wakefields.

A photoinjector with an output energy of 5 MeV produces the witness electron bunch, which is then accelerated to 10–20 MeV in a 1-m-long booster structure [8]. A 15-m-long transfer line [9] finally transports the bunch from the booster to the rubidium vapor source. The electron source can provide an electron bunch charge between 0.1 and 1 nC. The nominal normalized emittance of the electron beam is 2 mm · mrad.

_________
[*]livio.verra@cern.ch

We use beam-position monitors (BPMs) to measure the position of the proton and electron beams along the beam line and scintillating screens (BTVs) to measure their transverse bunch profiles [10]. Losses and radiation produced by the proton beam are monitored by proton beam loss monitors (PBLMs) positioned along the transfer line and the vapor source. Figure 1 shows a schematic of the beam transfer line and of the vapor source close to the vapor source entrance and the relevant diagnostics devices.

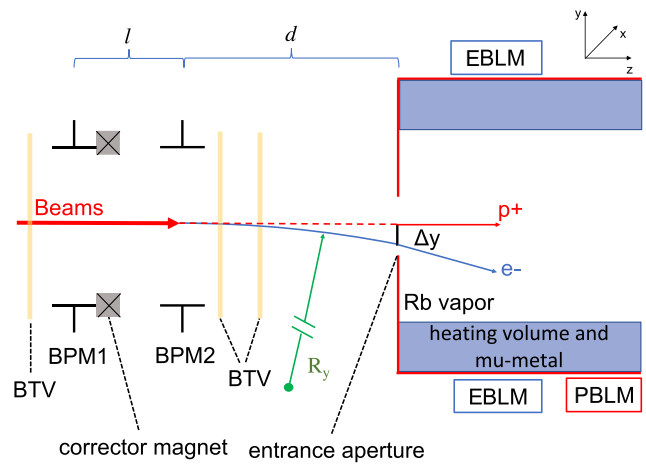

FIG. 1. Schematic of the proton and electron beam transfer line and vapor source close to the vapor source entrance. Main beam diagnostics devices are highlighted: beam-position monitors (BPMs), scintillating screens (BTVs), and electron and proton beam loss monitors (EBLMs and PBLMs, respectively). *l* and *d* are the distances between the two BPMs and between the last BPM and the vapor source entrance aperture, respectively. Beams are overlapped at the last two BPMs: The proton beam propagates essentially straight (red arrow); the electron beam trajectory (blue) is bent by Earth's magnetic field with radius of curvature $R_y$. $\Delta y$ is the deviation from the straight trajectory in the vertical plane. The drawing is not to scale.

### B. Experimental challenges

To inject and accelerate the electrons, we spatially and temporally overlap them with the plasma wakefields [11]. This means that the electron and proton beam trajectories have to cross within the plasma cylinder. To investigate the acceleration process and to characterize the wakefields, we want to inject the electron bunch at various locations downstream from the plasma entrance and at various angles with respect to the proton beam trajectory. We observe that the highest capture and acceleration efficiency occurs when the electron beam is injected ∼1 m downstream from the entrance. This is therefore the baseline setup for the acceleration experiment.

Because of the complexity of the vapor source, it was not possible to install any beam position or beam size diagnostics close to or along the plasma. Therefore, the last direct measurement of the electron beam is given by a scintillating screen positioned 0.8 m upstream from the entrance of the vapor source. Furthermore, during the acceleration experiment, no screen can be inserted in the beam line, because this would completely absorb the electron beam. This makes the alignment process for the injection extremely challenging due to the uncertainty on the electron transverse beam size at the injection point and due to the different effects of external magnetic fields on the two beam trajectories, given by the very different rigidity.

The rms transverse size $\sigma$ at the crossing point is one of the factors that contributes to the charge capture efficiency. Measuring the size near the crossing point is therefore important. Moreover, including the effect of Earth's magnetic field on the low-energy beam is crucial to precisely predict the electron beam trajectory only using information provided by BPMs.

In this article, we illustrate how we use the electron beam loss monitors (EBLMs) to measure the transverse beam size at the plasma entrance and infer it at the injection point. We also use this setup to align the proton-electron beam trajectories, by measuring the effect of Earth's magnetic field on the electron beam trajectory.

## II. MEASUREMENT SETUP

When electron and proton beams interact with the material surrounding the vapor source, they generate beam losses in the form of scattered and secondary particles. To detect these losses, we installed two EBLMs 1.5 m downstream of the source entrance aperture as shown in Fig. 1.

Each detector consists of two main parts, optically connected by a light guide: a scintillating material (EJ-200, a polyvinyltoluene-based plastic organic scintillator) and a photomultiplier tube (PMT) biased with a negative high voltage (∼kV). When particles cross the detector material, they deposit energy; part of this energy is converted to scintillating light that is transmitted to the PMT via the light guide. The PMT produces an amplified electric signal, read out by an oscilloscope. We control the amplification power of each detector independently, with the high voltages applied to the PMTs. These are chosen such that the detectors respond linearly to our range of deposited energies. The linearity of the system has been checked varying the charge of the incoming beam, with fixed trajectory, while measuring the loss signals [12]. The integral of the output signal is proportional to the charge produced by the PMT, i.e., to the deposited energy, and it is indicated as *counts*. In the following text, losses will be expressed in percentage with respect to the maximum counts value of each given dataset.

## III. MEASUREMENTS CONCEPT

As mentioned in Sec. I A, the vapor source has a 10-mm-diameter aperture in a 600-$\mu$m-thick aluminum foil for the rubidium vapor to exit the source. The thickness of the foil and the size of the aperture have been chosen according to the mechanical and thermal constraints of the vapor source, to minimize the radiation produced by the proton beam during the acceleration experiment and to allow for oblique external injection of the electron beam [11]. When beam particles hit the aluminum entrance foil, they produce secondary particles (scattered electrons and x rays) that deposit energy in the beam loss monitors. The thickness of the foil is sufficient to produce a high signal-to-noise ratio in the detectors when a fraction of the 18 MeV electron beam interacts with the material. The foil thickness and the distance between the foil and detector can, in principle, be adjusted to obtain a suitable signal. The loss signals are proportional to the amount of beam interacting with the material. Measuring these losses, we calculate the electron transverse beam size at the entrance aperture location and the deflection from the straight trajectory caused by Earth's magnetic field on the electron beam.

### A. Transverse beam size measurements

The goal of the measurement is to predict the rms transverse electron beam size $\sigma$ at the injection point, in order to improve the trajectory pointing precision and to estimate the charge capture efficiency. To effectively inject the witness bunch into the wakefields, its transverse size has to be comparable to the transverse extent of the plasma wakefields. This is given by the plasma skin depth $c/\omega_{pe}$, where $c$ is the speed of light and $\omega_{pe} = \sqrt{n_e e^2/\epsilon_0 m_e}$ is the plasma electron frequency ($n_e$ is the plasma electron density, $e$ is the elementary charge, $\epsilon_0$ is the vacuum permittivity, and $m_e$ is the electron mass) [13]. For a plasma electron density of $2 \times 10^{14}$ cm$^{-3}$, $c/\omega_{pe} \approx 0.4$ mm. We cannot directly measure the electron beam $\sigma$ at the injection point, as it is located ∼1 m downstream from the vapor source entrance. Therefore, we measure the beam size at the plasma entrance and estimate the size at the injection location, from beam optics.

We use the last corrector magnet in the beam line (see Fig. 1) to scan the electron beam position horizontally and vertically across the entrance aperture (examples of electron beam transverse positions at the entrance aperture are shown in Fig. 2) while recording the signals of the electron beam loss monitors. This beam scraper technique is a well-known and routinely used procedure in machine operation for beam collimation and aperture size measurements [14,15] and for transverse beam profile measurements [16]. Using the horizontal and vertical beam positions measured on BPM1 and BPM2, we reconstruct the horizontal and vertical $(x, y)$ position of the electron beam at the entrance location using a linear trajectory prediction:

$$(x, y) = \frac{(x_2 - x_1, y_2 - y_1)}{l} \cdot d + (x_2, y_2), \quad (1)$$

where $x_{1,2}$ and $y_{1,2}$ are the horizontal and vertical beam position measurements (offset from the center of the beam line) given by BPM1 and BPM2, respectively, $l$ is the distance between the two BPMs, and $d$ is the distance between BPM2 and the plasma entrance. Even though BPM1 is positioned upstream of the corrector magnet, we use its measurement as the beam position at the exit of the corrector, since the two instruments are only ∼9 cm apart and the position deviations at the exit of the magnet are small (< 0.05 mm). We also neglect the effect of Earth's magnetic field on the electron beam trajectory, as it gives a constant deflection (see Sec. III B) and is, thus, not relevant for beam size measurements. For each electron beam position at the aperture, we collect and average 30 measurements. The electron beam normalized emittance was measured to be ∼9 mm mrad with a quadrupolar scan at the exit of the electron source.

Figure 3 shows one side of the vertical and horizontal scans of the 200 pC electron bunch focused at the entrance aperture, measured by the detector positioned above the vapor source. We note that the minimum of the measured losses is around 5% (position at the entrance < 3.5 mm in Fig. 3), when the beam is centered on the aperture. We attribute this small, but nonzero, value to the non-Gaussian halo of particles around the Gaussian bunch. As soon as a significant number of beam particles hit the aluminum entrance foil, losses increase, reaching a maximum when they all interact with the iris (> 6 mm in Fig. 3).

Assuming that the transverse electron beam charge distribution is Gaussian [9], we can fit independently both rising ramps of each loss scan with an error function

$$\mathrm{erf}(x; \mu, \sigma) = \frac{1}{\sqrt{2\pi\sigma^2}} \int_0^x e^{-[(t-\mu)^2/2\sigma^2]} dt, \quad (2)$$

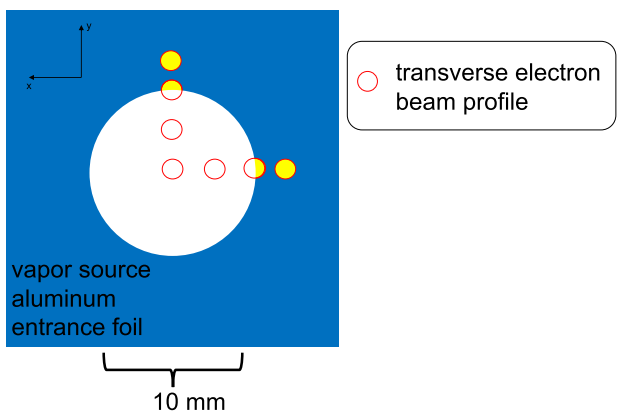

FIG. 2. Schematic drawing of the vapor source entrance. Examples of electron beam transverse positions at the entrance aperture during the horizontal and vertical scans are shown. The yellow areas mark the fraction of the beam interacting with the material, i.e., beam loss. The drawing is not to scale.

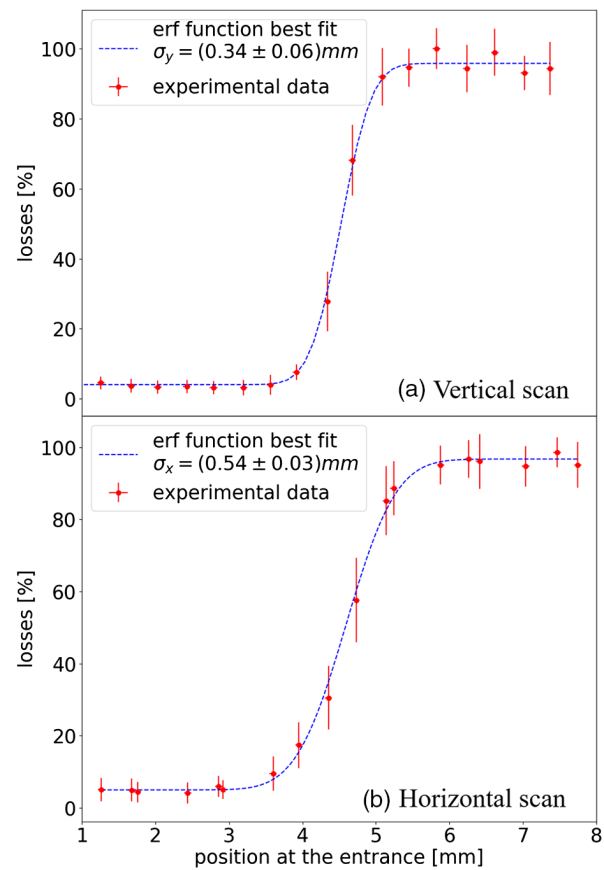

FIG. 3. Loss signals (red dots) measured as a function of the vertical (a) and horizontal (b) position at the vapor source entrance [calculated with Eq. (1)]. Every point is the mean value of 30 measurements; error bars are the standard deviation of the distribution for each point. Each plot is fitted with an error function according to Eq. (2) (blue dashed lines). For these measurements, the 200 pC electron beam is focused at the vapor source entrance (measurement location).

TABLE I. Results of the electron beam scan of the entrance aperture for different focal point locations. Every scan gives two transverse beam size values $\sigma_{\text{fit}}$, one for each side of the beam loss signal. The final value of $\sigma$ is obtained as the mean of the two measurements for each scan; the error is calculated propagating the statistical error.

| Focal point location | Dimension | $\sigma_{\text{fit}}$ [mm] | | $\sigma$ [mm] |
|---|---|---|---|---|
| Entrance aperture | Vertical | $0.28 \pm 0.03$ | $0.34 \pm 0.06$ | $0.31 \pm 0.03$ |
| | Horizontal | $0.53 \pm 0.04$ | $0.54 \pm 0.03$ | $0.54 \pm 0.03$ |
| 1 m downstream from the entrance | Vertical | $0.52 \pm 0.05$ | $0.4 \pm 0.2$ | $0.5 \pm 0.1$ |
| | Horizontal | $1.13 \pm 0.06$ | $0.98 \pm 0.08$ | $1.06 \pm 0.05$ |

where $\mu$ is the position of the center and $\sigma$ the rms of the Gaussian distribution. Every loss scan produces two values of the beam transverse size $\sigma_{\text{fit}}$ (one for each side).

According to the beam line optics [9], we also focus the beam 1 m downstream from the entrance aperture and repeat the measurement, that is, the optical configuration used during the injection experiment. In Table I, we give the resulting $\sigma_{\text{fit}}$ values for the vertical and horizontal scans for the different optics. The error on $\sigma_{\text{fit}}$ is given by the fit covariance matrix and, therefore, quantifies the goodness of the fit.

We note that the two values of $\sigma_{\text{fit}}$ for each scan agree with each other. The final values are calculated as the mean of the two measurements for each scan, as given in Table I; the errors on the final values are calculated propagating the statistical error on the single measurement. When the beam is focused at the entrance, $(\sigma_x, \sigma_y) = (0.54 \pm 0.03, 0.31 \pm 0.03)$ mm; when it is focused 1 m downstream, $(\sigma_x, \sigma_y) = (1.06 \pm 0.05, 0.5 \pm 0.1)$ mm. The measured vertical transverse beam size at the waist (beam focused at the entrance) is slightly larger but still consistent, within 2 times the statistical error, with the nominal value (0.25 mm) [9]. The horizontal $\sigma$ is measured to be larger than the vertical one in both optical settings; the beam is, therefore, not round as expected from the design. The difference is attributed to the dispersion $D$ in the horizontal plane [17], that is minimized at the beam waist but never fully compensated.

The measurements performed focusing the beam 1 m downstream from the vapor source entrance provide a value of the transverse beam size 1 m upstream from the waist position. Therefore, we calculate the beam size at the waist $\sigma_0$ according to linear Gaussian beam optics: $\sigma(z) = \sqrt{(\sigma_0^2 + z^2 \epsilon_g^2/\sigma_0^2) + (D\delta p/p)^2}$ (where $z = 1$ m, $\sigma$ is the transverse beam size obtained from the measurements, $\epsilon_g$ is the geometric emittance, and $\delta p/p \sim 0.5\%$ is the momentum spread [9]). Thus, this measurement allowed us to predict the beam transverse size at the injection point as $(\sigma_x, \sigma_y) = (0.60 \pm 0.06, 0.3 \pm 0.1)$ mm (the errors are calculated propagating the statistical errors obtained above). For a plasma electron density $n_e = 2 \times 10^{14}$ cm$^{-3}$, we are thus confident that a significant fraction of the bunch is injected into the wakefields (when the beam trajectory is properly set to cross the wakefields).

To further test this measurement concept, we also measure the transverse beam size of the well-characterized SPS "pilot" proton bunch ($\epsilon_N \sim 1$ mm mrad, bunch population $= 10^{10}$ particles). We directly see from Fig. 4 that the slope of the proton bunch (blue curve, rise ramps in the $[-6, -4]$ and $[4,6]$ mm ranges) is steeper than that of the electron beam signal (red curve, rise ramps in the $[-4, -2]$ and $[5,7]$ mm ranges). This indicates that the proton beam transverse size is smaller than the electron beam one. With the same fit procedure described above, we measure it to be $\sigma = (0.12 \pm 0.02)$ mm. It is in good agreement with the expected value $(0.10 \pm 0.01)$ mm: This is calculated measuring the proton beam $\sigma$ with foils emitting optical transition radiation upstream and downstream from the vapor source and the beam emittance in the SPS.

### B. Electron beam deflection from Earth's magnetic field

The externally injected electrons have a low energy ($\sim 18$ MeV), and the transfer beam line is not shielded from external magnetic fields. Earth's magnetic field $B$ in the experimental area was measured during the installation campaign to be [18] $B_{(x,y)} \sim (0.2, 0.4)$ G (with a $\pm 15\%$ uncertainty), corresponding to a Larmor radius $R_{(x,y)} = \beta\gamma m_e c/eB_{(y,x)} \sim (1.5, 3)$ km. In particular, the beam trajectory between the last magnetic element and the entrance of the vapor source (more than 3 m away) cannot be approximated as straight, since Earth's magnetic field bends the beam onto a circular trajectory. The vapor source is shielded with mu metal, so that the electron beam trajectory is straight, once injected into it.

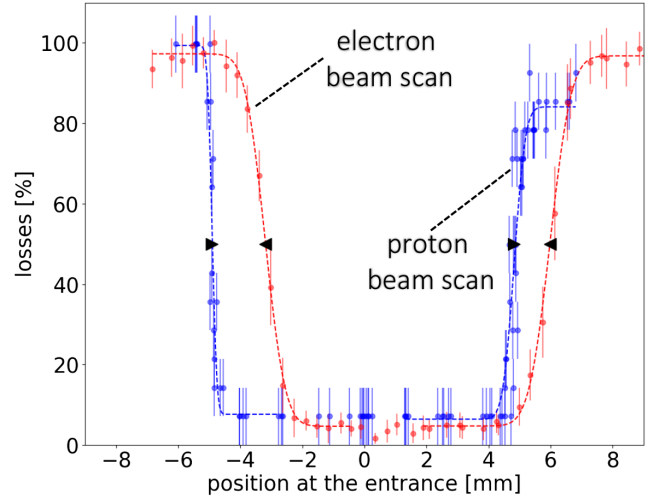

FIG. 4. Proton (blue dots) and electron (red dots) beam losses as a function of the horizontal position at the vapor source entrance aperture. Dashed lines are the error function fits, the black triangles the centers of the rising ramps.

We estimate the deviation from straight trajectory as (see Fig. 1) [12]

$$\Delta x, \Delta y \sim d \sin\left(\frac{1}{2}\frac{d}{R_{x,y}}\right). \quad (3)$$

The beam position at the vapor source entrance is predicted to be different from that given by a straight line trajectory by $\Delta x \sim -1.3$ mm (to the right in the horizontal plane) and $\Delta y \sim -0.66$ mm (down in the vertical plane), with an uncertainty of $\pm 15\%$.

Since the last BTV is too close to the BPMs to resolve the trajectory deviation, and no beam size or position instrument can be installed at the plasma entrance, it is not possible to directly measure this electron beam trajectory deflection. Thus, we developed an indirect measurement technique that uses both the proton and electron beam loss monitors and the vapor source entrance aperture as follows:

1. *Proton beam scan to establish position references.*—While recording the loss signals from the proton beam loss monitor (positioned on the right-hand side of the vapor source and downstream from the entrance aperture), we scan (horizontally and vertically) the proton beam position over the entrance aperture by shifting the beam parallel to its nominal trajectory (see the blue dots in the horizontal scan of Fig. 4). Note that losses on the negative side (right-hand side) are higher than on the positive side because of the position of the detector. We fit both rise ramps with the error functions [Eq. (2)] and define the position of the entrance aperture edge in the two transverse dimensions as the $\mu$ values of the rising ramps (black triangles pointing right in Fig. 4). A straight trajectory prediction of the proton beam trajectory is justified, as the effect of Earth's magnetic field on the 400 GeV/c proton bunch is smaller in amplitude than on the electron bunch by a factor $p_{p+}/p_{e-} = 2.6 \times 10^4$, where $p_{p+,e-}$ is the momentum of the proton and electron beam, respectively. Using the loss scans, we align the proton beam position on the center of the entrance aperture, and we take a trajectory reference on two scintillating screens upstream from the vapor source.

2. *Electron beam scan.*—After aligning the electron beam onto the proton reference trajectory at the scintillating screens (and, therefore, including in the measurement offset readings of the BPMs), we scan horizontally and vertically the electron beam position over the aperture while recording the EBLM loss signals. Then, we compute the beam position at the iris using Eq. (1) (red dots in Fig. 4) and fit the ramps with error functions [Eq. (2)], obtaining the $\mu$ values (black triangle pointing left in the plot) as the centers of the ramps. The error on $\mu$ is provided by the covariance matrix of the fit.

3. *Comparison of loss signals.*—As shown in Fig. 4, the proton and electron beams loss distributions do not overlap in space because of the effect of Earth's magnetic field on the electron beam trajectory. Thus, we determine the deflection $(\Delta x, \Delta y) = (\mu_{p^+} - \mu_{e^-})_{x,y}$, where $\mu_{p^+,e^-}$ are the centers of the rising ramps for the proton and electron scans, respectively. As the $\sigma$ of the two beams are different, we obtain two values of the deflection for each plane (see right- and left-hand sides of the scans in Fig. 4). We use the mean of the two as a final estimate of the deflection.

The measured values are $\Delta x = (-1.44 \pm 0.03)$ mm (to the right in the horizontal plane) and $\Delta y = (-0.55 \pm 0.03)$ mm (down in the vertical plane). The measurements agree with the calculations discussed above ($\Delta x \sim -1.3$ mm, $\Delta y \sim -0.66$ mm) $\pm 15\%$. This allows us to reach true electron-proton beam crossing at the plasma entrance. Correcting the electron beam trajectory upstream, we could also make the two beam tangent at their crossing point, aligning the position and angle. This trajectory is then used as a reference for injection during the acceleration experiment.

## IV. CONCLUSIONS

Using the electron beam loss monitor setup, we conduct measurements on the AWAKE electron beam. Measuring losses at the vapor source entrance aperture when the beams are made to hit the aperture, we measure the transverse beam size of the electron beam for two different magnetic optic settings. The results agree with the optical model of the beam line. This measurement has been essential for the electron beam line commissioning and for the external electron injection experiment, since no other beam transverse size diagnostics is available at that location: The EBLM system provides the closest information about the electron beam size and position to the injection point ($\sim 1$ m downstream from the aperture).

Using the same technique, we measure the deflection of the low-energy ($\sim 18$ MeV) electron beam trajectory, after the last corrector magnet, caused by Earth's magnetic field. We use this information to correct the electron beam trajectory in order to make it cross with the proton bunch trajectory at the desired location.

We note that this beam loss method is applicable when the beam is smaller than the entrance aperture but larger than the uncertainty on the transverse position. This method could be used in advanced accelerator experiments, when the electron beam for external injection into wakefields must be aligned onto the center of a capillary discharge or gas cell. These have, in general, rather small apertures ($\leq 1$ mm) and the beam must be aligned in position and angle.